# Transformation of the trapped flux in a SC disc under electromagnetic exposure

V. V. Chabanenko[1], I. Abaloszewa[2], V. F. Rusakov[3], O. I. Kuchuk[2], O. M. Chumak[2], A. Nabiałek[2], A. Abaloszew[2], A. Filippov[4], R. Puźniak[2]

[1] O. Galkin Donetsk Institute for Physics and Engineering, National Academy of Science, Kyiv, Ukraine
[2] Institute of Physics, Polish Academy of Sciences, Warsaw, Poland
[3] National Technical University of Ukraine "Igor Sikorsky Kyiv Polytechnic Institute", Kyiv, Ukraine
[4] Christian-Albrechts-Universität zu Kiel, Kiel, Germany

**Abstract**

Superconducting (SC) permanent magnets with trapped magnetic flux are used in technical devices (motors, generators, etc.). These magnets endure repeated magnetic "shocks" during operation, which may influence their performance. In this work, we investigated the dynamic behavior of magnetic induction in the trapped flux in a SC disk when exposed to stepwise changes in the external magnetic field, simulating these operational shocks. Our results reveal a direct correlation between the stepwise changes in the magnetic field and the trapped flux response, with each increase or decrease in the field, inducing a corresponding 40–50% change in trapped flux for a 600 G field step at temperature of 5 K. The magnitude of these changes depends on the external parameters and their dynamics could lead to additional energy dissipation and potential heating, which may affect the reliability of SC magnets in applications. A scaling analysis of the induction flux profiles, revealing roughness exponents in the range of 0.435 to 0.475 was performed as well, and we determined the Hausdorff dimension of the surface structure.

## 1. Introduction

Superconducting (SC) materials with trapped magnetic flux are widely applied as permanent magnets in various equipment, such as motors, generators, and other devices [1-7]. The magnitude of the induction of permanent magnets in bulk of high-temperature superconductors is impressive: $B = 17.6$ T at $T = 26$ K [8-12]. During operation, such permanent magnets are subjected to continuous magnetic "impacts" of varying magnitude and frequency, for example, the interaction of stator and rotor magnets in an electro-motor, generator, magnetic bearing and magnetic pulse rings in the field of rotating machinery [13].

In high-temperature superconductors (HTSCs), magnetic shocks (along with thermomagnetic avalanches) trigger mechanical stress pulses in the material through the magnetostriction mechanism [14, 15]. This can lead to the mobility of impurities and defects, such as pinning centers, and may potentially cause material cracking [16, 17], ultimately transforming the single-peak magnetic induction [12, 18] into a multi-peak configuration [19], reducing the effectiveness of these magnets in devices.

In a generator [4-6], the movement of a SC disk magnet relative to a closed-loop coil, where an electromotive force (EMF) is induced, is accompanied by magnetic "rocking" of the trapped flux inside the magnet. As the magnet approaches the coil, electromagnetic interaction generates an oppositely directed magnetic field in the coil, reducing the trapped flux in the magnet. This field reaches its maximum when the disk fully overlaps the coil's cross-section. Upon further movement, the induced magnetic field reverses direction, effectively increasing the trapped flux in the magnet. Therefore, the magnetic interaction between the magnet and the generator coil can deform the induction distribution of the moving SC magnet. The induced field strength strongly depends on the magnet's rotation speed.

In our experiment, the stepwise increase/decrease of the external magnetic field in which the SC disk is located is, to a certain extent, an analogue of the process of forced dynamics of the magnet induction distribution, that occurs when it rotates relative to the generator coils [6].

Therefore, understanding the stability of the critical state of superconductors with trapped flux and their response to stepwise external magnetic field changes (similar to magnetic "shocks") is crucial for optimizing their operation. In this connection, the study of the sensitivity of the critical state of superconducting devices to various types of electromagnetic interference, for example, in the form of an alternating current magnetic field superimposed on a linearly increasing field [20], is also of interest.

Using magneto-optical (MO) imaging [21-23], we investigated the effects of stepwise changes in the external magnetic field on the flux structure in the critical state of an NbTi disk (a hard superconductor) with well-known properties [24-26]. The technological process of alloy processing includes the transformation of the microstructure of pinning centers (thermomechanical process – extrusion and subsequent heat treatment [27]) to improve the current-carrying properties of the material. This has made the NbTi one of the most widely applied superconductors.

To enhance the efficiency of SC magnets in applications, reducing losses is essential. In rotating magnets (e.g., generators), certain losses are attributed to both hysteresis losses within the superconductor itself and to the dissipation of eddy currents in surrounding conductive materials [28, 29].

In our study, we captured magnetic induction distribution patterns, which reflected the effects of dynamic external magnetic field perturbations on the critical state of the NbTi disk. Analysis of the experimental data revealed that a stepwise decrease (increase) in the external magnetic field leads to a corresponding increase (decrease) in the trapped flux in the NbTi disk; the magnitude of these changes of induction is 40-50% at $T = 5$ K. The induced flux motion can lead to additional energy dissipation in technical products and impose operational limitations.

Thus, the dynamic changes in the superconducting magnet's flux structure, driven by regular external magnetic shocks (such as in generators), result in dynamic resistance and losses [30]. This process, along with hysteresis and Joule losses, can affect change in the temperature of the device, impacting its operational stability and lifespan.

Due to demagnetization field, limited-volume magnets contain regions of opposite magnetization [31, 32], which are separated by well-defined boundaries, as demonstrated in disk-shaped SC magnets [33]. Experimental data for $YBa_2Cu_3O_{7-x}$ crystals [32] show that the induction profiles between regions of opposite polarity feature a single oscillation with a sharp magnetic wall at the midpoint of the transition. This wall behaves like a viscous medium (akin to "honey") and moves towards the SC center as the magnetic field increases. The dynamics of such a magnetic structure under stepwise external magnetic field changes, caused by vortex/antivortex annihilation, is also a dissipative process.

We studied how the pattern of magnetic field penetration changes under the influence of well-established technological processes (extrusion and long-term heat treatment) [26] used to create NbTi superconductors with high current-carrying capacities. Magneto-optical imaging revealed the presence of inhomogeneities and rough induction profiles in the alloy. The distribution of induction inside the superconductor exhibits a complex 3D ridge-like topology, and the boundary between the Meissner and mixed states displays a bent relief structure. Scaling analysis of the flux profiles' power function [34, 35] behind the front allowed us to determine their roughness index and Hausdorff dimension. We have carried out a comprehensive study of the induction profiles at the boundary of the Meissner state and behind the flux front at various levels.

This study proposes to use the MO method to assess the homogeneity of pinning centers across the entire volume of a superconducting alloy through alternating field remagnetization. This procedure forms magnetic domain boundaries with opposite field orientations throughout the SC volume. Analysis of such domain boundary structures, using roughness coefficients, for instance, allows us to characterize qualitatively the degree of pinning center distribution inhomogeneity in the material.

The presented results clearly demonstrate the applicability of a relatively simple MO method to improve the technology of obtaining SC materials, tracking the transformations of the structure of pinning centers and, accordingly, changes in the conductive properties.

## 2. Experimental methods and sample

We employed the magneto-optical technique to observe magnetic flux penetration and to map the induction distribution in the superconducting NbTi disk under stepwise changes in the external magnetic field. It is known [36, 37] that in the MO method with crossed polarizers, the normal component of magnetic induction, $B_z$, relative to the sample surface is measured. In MO images, brighter areas correspond to regions with a higher normal induction component ($B_z$), while darker areas indicate regions with lower induction. The MO method with crossed polarizers allowed us to distinguish the direction of the penetrating magnetic flux in the images. In this case, brighter image areas correspond to one flux direction, and darker areas correspond to the opposite direction. Both setups were utilized in our experiments to obtain induction profiles for analyzing the dynamic impact of external magnetic fields on the trapped magnetic flux.

Optical microscopy and digital images analysis of magnetic field induction in the superconductor, being effective for visualizing structures ranging from several millimeters down to microns in size [36], allowed us to estimate the scale of structural objects that affect magnetic flux pinning at the initial stage of material processing. This technique operates based on the Faraday rotation principle, using a magnetic Bi-substituted yttrium iron garnet (YIG) indicator, placed directly on the superconducting sample within a continuous-flow cryostat (temperature range: 4–300 K) and under a low-field magnet. The plane of polarized light reflected by an aluminum mirror situated at the bottom of the YIG indicator rotates, with the rotation being proportional to the local magnetic induction. This produces an image showing the magnetic flux distribution across the sample.

Figure 1(a) shows a photograph of the assembly, with the NbTi superconducting disk affixed to a copper base and two YIG indicators on its surface. Figure 1(b) presents a magneto-optical image viewed through a microscope eyepiece, showing the magnetic field structure ($B_{ext} = 600$ G) penetrating the disk's central region, followed by the external field being turned off at $T = 5.9$ K.

The initial cylindrical NbTi rod, consisting of a 50% alloy with a diameter of 50 mm, was extruded to a diameter of 15 mm at 750 ºC. Hysteresis loop measurements of a sample plate of the extruded NbTi material provided the following alloy parameters: $B_{c2}(T = 4.2$ K$) = 11.1 \times 10^4$ G, $j_c(T = 4.2$ K; $B_{ext} = 0$ G$) = 4 \times 10^9$ A·m$^{-2}$; $B_{c1} = 10$ G at 5–6 K.

The disk-shaped sample, with a thickness of $h = 0.1$ mm and a diameter of $D = 12$ mm, was cut from the extruded rod after hydroextrusion, and its damaged surface layer was removed by grinding. One side of the disk was polished to a mirror finish to reflect the incident polarized light. After initial studies, the extruded material was subjected to annealing. The annealing parameters of such NbTi materials have been studied well already [26, 27]. As detailed there, the superconducting alloy with the highest critical current density was produced using various strain regimes and multiple long-duration heat treatments. According to Ref. [27], the optimal combination includes three heat treatments, each lasting 80 hours at 420 ºC. In our case, one stage only of such processing was applied and the changes in the flux front structure in the superconducting alloy were observed.

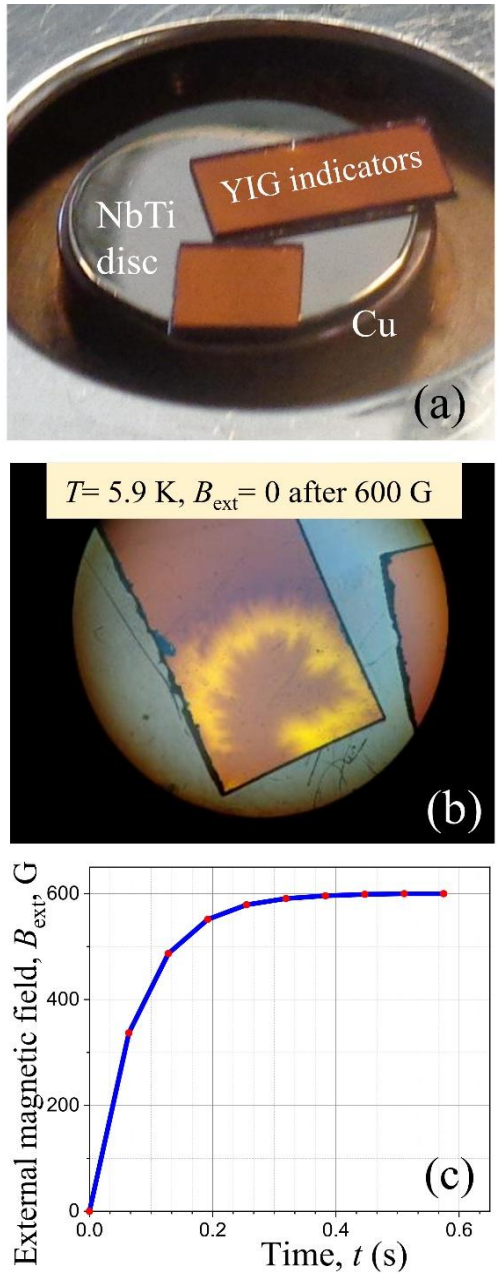

Figure 1. (a) – Photograph of an assembly of NbTi superconducting disk, with a thickness of $h = 0.1$ mm and a diameter = 12 mm, glued to a copper (Cu) base; there are two plates of magneto-active material (YIG indicators) on the surface of disk; (b) – view in the microscope eyepiece: magneto-optical image of the structure of the penetrating magnetic field into the central part of the disk (NbTi after extrusion); external magnetic field $B_{ext} = 0$ after 600 G, $T = 5.9$ K; (c) – typical step of external magnetic field dependence $\Delta B_{ext}$ vs $t$.

The magnetic field was applied perpendicular to the disk surface. The flux front in the superconductor was manipulated using stepwise changes in the external field, $\Delta B_{ext}$, with steps of 50 and 100 G, up to a maximum of 600 G. Additionally, stepwise field changes of $\Delta B_{ext} = \pm 600$ G were applied. The typical dependence of stepwise magnetic field changes, $\Delta B_{ext}$ vs. $t$, is presented in figure 1(c). The rate of field change was fixed to $dB_{ext}/dt \leq 0.5$ T/s, which is an order of magnitude slower than the rates typically used in pulsed magnetization of hard superconductors [38]. After each stepwise change in the external magnetic field, the resulting induction distribution was recorded via magneto-optical imaging. Measurements were conducted in gaseous helium at temperatures ranging from $T = 5$ to 8 K.

The disk's magnetization, $M$, was determined as the differential signal between two Hall probes. One probe was placed on the surface of the sample's center, measuring the local magnetic field $B_{surf}$, while the second probe measured the external magnetic field $B_{ext}$.

## 3. Magnetic flux front dynamics and distribution of induction: results of the experiment

In the experiment, the sample was cooled to a temperature below $T_c$ in zero magnetic field. Once the external field was applied, the resulting induction distribution in the superconductor was recorded. The penetrating magnetic flux vortices interacted with the microstructure of the material, which include defects, impurities, and grain boundaries. These features, through various mechanisms, contribute to the pinning of the magnetic flux. Consequently, the magnetic field induction distribution within the superconductor reflects the structure of the pinning centers in the material.

The following section presents the changes in magnetic induction within the superconductor after undergoing a thermomechanical process [27] – hydroextrusion followed by heat treatment.

### 3.1 Magneto-optical visualization of the mechano-thermal effect on the pinning structure

The induction distribution at specific external magnetic field steps ($\Delta B_{ext} = 50, 100, 200, 400$ and 600 G) at temperature of 6 K are shown in figure 2. The left column (a–e) represents the data for NbTi alloy after extrusion, while the right column (g–k) shows the sample after heat treatment of the extruded material. In the images, the Meissner state of the superconductor appears dark, while the incoming magnetic flux is represented by bright areas. The yellow line marks the flux front, representing the boundary between the critical state and the Meissner state ($B = 0$). From the comparison of the field penetrations in the left and right columns, it is clearly seen that the shielding of the external magnetic field increases after annealing the SC alloy. Significant changes are also observed in the magnetic field induction distribution.

Figure 2(f, l) shows the magnetic flux front lines at the Meissner level (yellow lines on the upper views) for two states of the alloy: extruded (f) and annealed (l). It is evident from figure 2(l) that the boundary of the magnetic flux penetration into extruded superconductor exhibits a more complex, rough structure: two scales of roughness are evident in the flux front lines across all external field values. One scale represents large, non-periodic flux front variations amplitude, while the other consists of smaller ripples that continuously cover the flux front lines. During extrusion, the crystal structure of the material undergoes significant deformation, resulting in the appearance of large-scale structural defects that serve as major pinning centers and lead to coarse roughness.

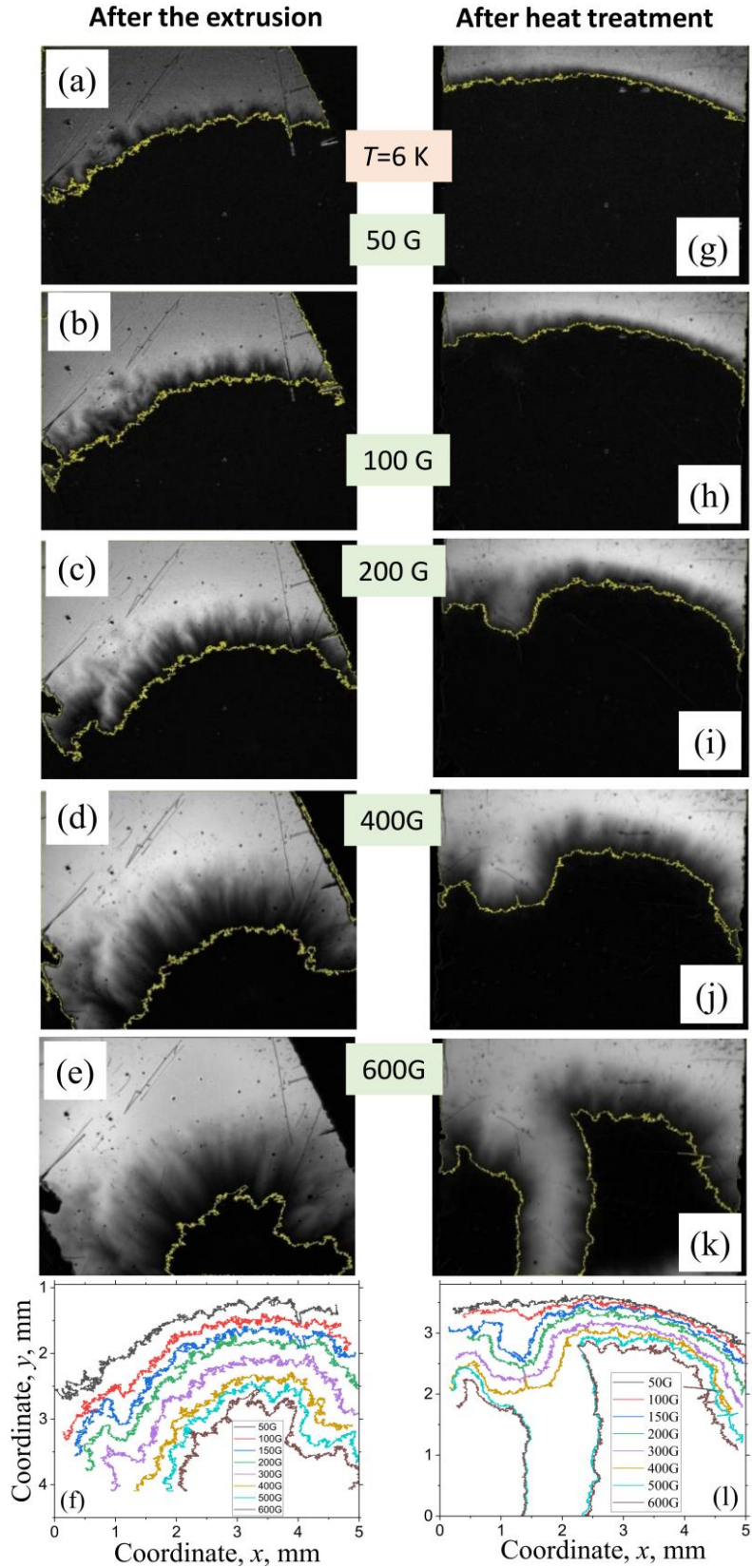

Figure 2. (a-e) and (g-k) - Flux patterns at magnetic field penetration; bright and dark areas correspond to the trapped flux and Meissner phases. Yellow lines are profiles of flux front on the Meissner level in various fields: $\Delta B_{ext}$ = 50, 100, 200, 400 and 600 G; (f, l) – the structure of flux front lines on the Meissner level. Left column (a-f) – NbTi alloy after extrusion and right column (g - l) - after heat treatment; NbTi disc with diameter of 12 mm and the thickness of 0.1 mm, $T$ = 6 K.

When comparing the distribution of induction in the left and right columns of figure 2, it becomes clear that after annealing, the structure of the pinning centers in the alloy is transformed to a more uniform with a reduction in the size of the pinning centers. As a result the flux front becomes smoother with the roughness finer in scale. This is particularly evident in the flux front lines in low magnetic fields, where for a field of 50 G (figure 2(g, l)), the large roughness features are almost entirely absent. Here, the front is represented only by small roughnesses. As the magnetic field increases, larger-scale roughness reappears, with an increase in amplitude (figure 2(k, l)).

As follows from figure 2(f, l), more efficient pinning of vortices leads to decrease of the depth of flux penetration into the annealed sample by 2–3 times, which indicates a 2–3-fold increase in the critical current density. The increase in critical current due to heat treatment introduces an instability in the critical state, resulting in a magnetic flux jump (see the flux finger in figure 2(k)). The flux breakthrough occurred in a weak spot – a small gate area visible in figure 2(i, g). In this region, the superconductor's current lines form semicircles with a small radius, creating a localized magnetic induction hill. This induction hill becomes the point of instability, initiating a flux avalanche. It should be noted that in the avalanche finger region (figure 2(l)), the flux front practically did not shift with an increase in the external field from 500 to 600 G, compared to the movement of the front in the region of the critical state belt.

The detailed structure of the magnetic field induction profiles in the superconductor will be analyzed later (chapter 3.4) using roughness exponents derived from power functions.

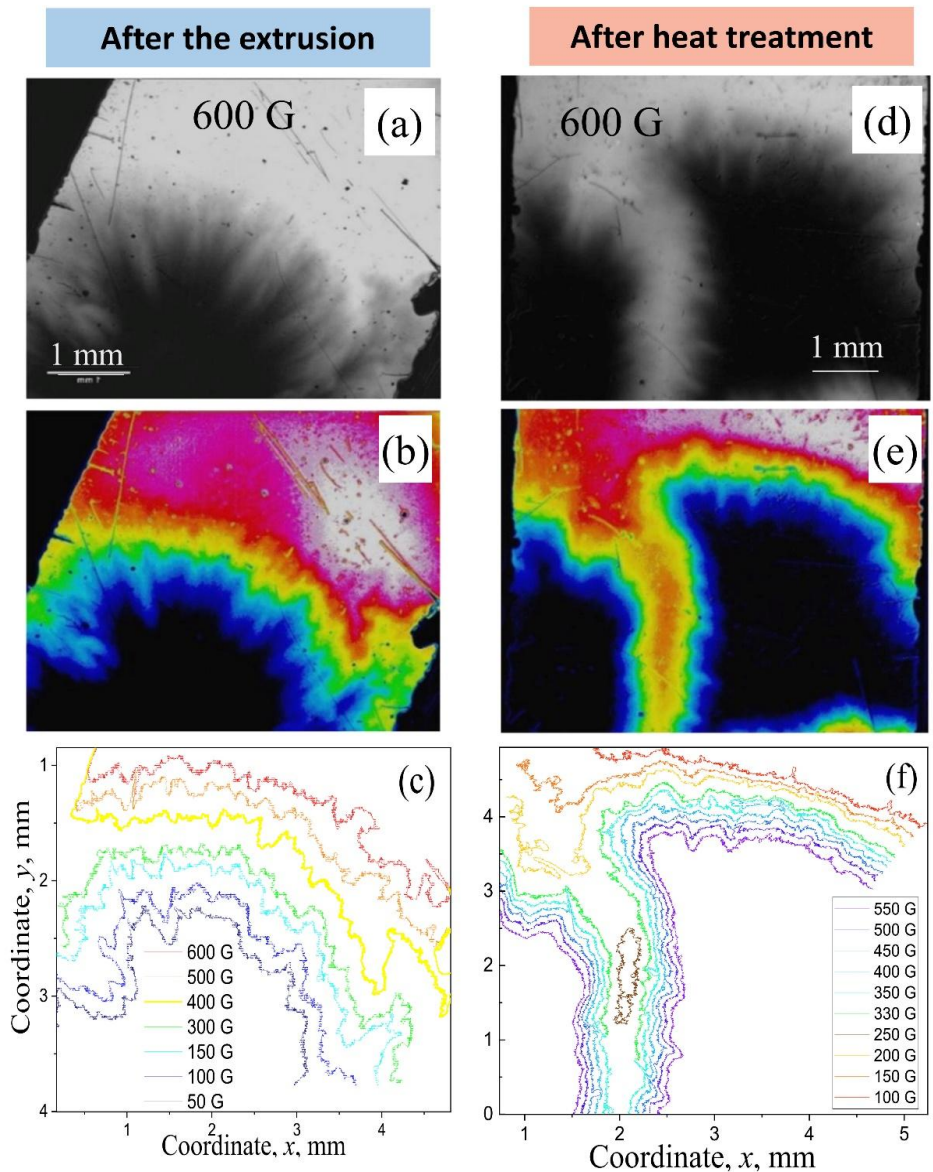

Figure 3. (a, b) and (d, e) - Penetration fronts of magnetic induction into NbTi disc at $B_{ext}$ = 600 G, $T$ = 5.6 K: (a, b) - after the extrusion and (d, e) - after the heat treatment. (c, f) - The induction contours in the area of flux penetration on various levels of induction $B$: (c) – 50, 100, 150, 300, 400, 500 and 600 G and (f) - 100, 150, 200, 250, 330, 350, 400, 450, 500 and 550 G.

Figure 3 shows the structure of magnetic induction of penetration front for an external field $\Delta B_{ext}$ = 600 G at $T$ = 5.6 K in the superconducting alloy after extrusion (a) and after annealing (d), with (b) and (e) displaying a color representation of these data. The iso-induction contours in the flux contours in the flux penetration region at various induction levels are presented on (c) for $B_{ext}$ = 50, 100, 150, 300, 400, 500 and 600 G and on (f) for $B_{ext}$ =100, 150, 200, 250, 330, 350, 400,

450, 500 and 550 G. The value of magnetic induction in the image was determined on the base of brightness intensity. An analysis of the iso-induction contour will be discussed later. As it follows from figure 3(c, f), the induction contours in the area of flux penetration becomes smoother for various levels of induction, and the roughness of it after annealing is finer as well as on the structures of flux front on the Meissner level (figure 2(f, l)).

In the central part of the finger, a maximum of the incoming flux is formed (figure 3(f), which is clearly visible in the color image (figure 3(e)). Such flux peaks are characteristic for thermomagnetic avalanches in other materials as well [39, 40]. Let's examine the small-scale roughness of the flux front in more detail. These features are likely associated with the primary pinning centers in the material, which may include α-Ti phase precipitates and structural defects that became finer after annealing [27].

Figure 4(a) presents the magneto-optic image of the magnetic flux in a 600 G field, recorded for the superconducting disk after heat treatment at $T = 5.5$ K. In figure 4(b), with the contrast significantly enhanced, small regions of magnetic inhomogeneities near the Meissner state boundary are clearly visible. Iso-induction lines at the level of several gausses, highlighting this fine-scale magnetic inhomogeneity, are shown in figure 4(c). The details of the iso-induction line structure are presented in figure 4(d), which shows one of the large magnetic domains located at the intersection of two orthogonal lines.

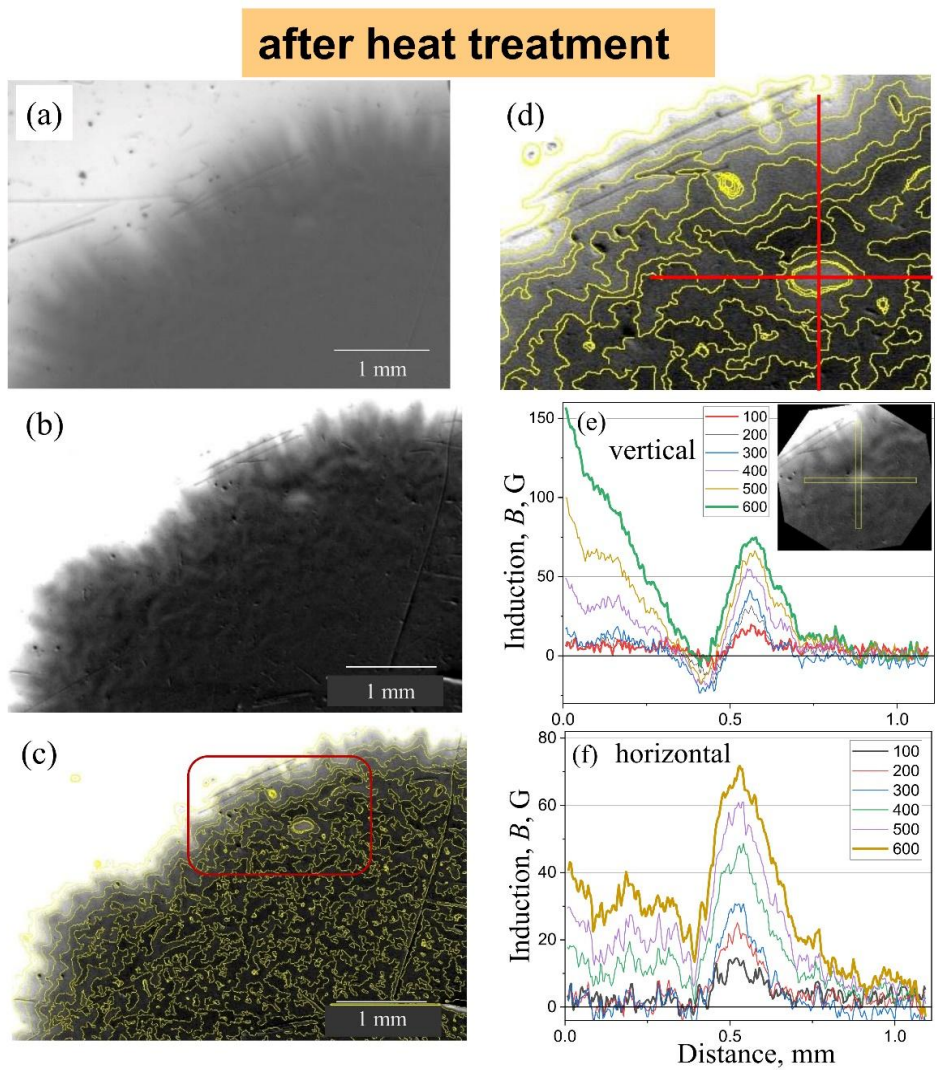

Figure 4. Flux patterns of magnetic field (600 G) penetration: (a) – magneto-optical image, (b) – the same with increased contrast and (c) – contours of constant induction on the level of several gausses; (d) – the details of induction structure in a rectangle on (c); (e) and (f) – profiles of induction around the grain along orthogonal directions, presented on (d) and on insert to (e), in various fields $B_{ext} = 100, 200, 300, 400, 600$ G; NbTi disc after heat treatment; $T = 5.5$ K.

The magnetic induction distribution around magnetic domain along the two indicated directions, with a discrete increase in the external field step from $\Delta B_{ext} = 100$ G up to of 600 G, is presented in figure 4(d). The induction distribution in figure 4(e, f) clearly demonstrates that the pinning centers on the surface of the superconducting disk trap a measurable amount of flux even at an external field of 100 G. The magnitude of the trapped flux in the magnetic domain (figure 4(e, f)) increases almost linearly with the rising external magnetic field. The size of the large flux-trapping domain presented in figure 4(d) is approximately equal to $0.2 \times 0.1$ mm².

It is important to note that the normal component of the external field at the disk surface is very small (figure 5 taken from Ref. [33]). According to the calculated distribution of induction in the disk [33], the field lines slide over the sample's surface towards the disk's center (figure 5), remaining nearly parallel to the surface. The portion of full transparency for the magnetic field through the disk's thickness, from its edge to the center, is around 15–20%, depending on the ratio between the external field induction and the field of complete penetration.

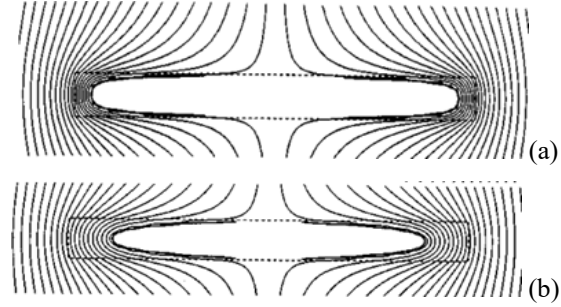

Figure 5. Magnetic field lines during flux penetration into a thin disk at $B_{ext}/B_p$ = 0.1 (a), 0.2 (b); $B_p$ is full penetration field. Figure was taken from Ref. [33].

*3.2 Dynamic changes in the flux structure with stepwise changes in the external field*

The magnitude of the trapped flux hill in the finger decreases when the field is increased from 500 to 600 G, as it is shown in figure 6(d). In the black-and-white induction representation, the flux hill at 600 G appears darker than that at 500 G. This reduction in the flux hill is even more evident in the corresponding color images (figure 6(c*, d*)).

The reduction in the flux hill's induction level is equal to about 5% only, despite a 20% increase in the external magnetic field from 500 to 600 G. The flux hill reacts more strongly when the external field is switched off at 600 G (i.e., for $\Delta B_{ext}$ = 600 G). The result is shown in figure 6(e, e*), where the induction peak significantly increases in amplitude. Thus, pulses of decreasing or increasing magnetic fields result in a corresponding increase or decrease in the magnitude of the trapped flux.

Switching the magnetic field in the opposite (-600 G) direction (figure 6(f, f*)) and subsequently turning it off (figure 6(g, g*)), differs from the previous process with switching off +600 G in that it includes an additional annihilation process of the previously trapped flux. As a result of additional heating, the field penetrates deeper into the sample (figures 6(f, f*)) than in the previous case (figures 6(d, d*)). However, the previously discussed dynamic effects are also observed in this case. Let us analyze the changes in the induction distribution in the SC disk due to the stepwise dynamic effects of the external field $\Delta B_{ext}$ = ±600 G. Such an analysis can be useful because real bulk superconductors often exhibit non-uniform parameters across their volume, resulting in magnets with heterogeneous induction distribution [4, 5].

Figure 7 shows the induction distributions at $T$ = 5.3 K in the longitudinal (a and b) and transverse (c, d, and e) directions after the following manipulations: switching on a 500 G field, increasing it to 600 G, and then turning it off. Processing of the data is identical to that presented in figures 6(c*, d* and e*). Arrows marked as 1 and 2 in figure 7(e*) indicate the directions along which the longitudinal induction profile analysis was conducted in $B_{ext}$ = 0 after 600 G. (profile 1 and profile 2). Profile 1 crosses the flux hill and the critical state region along the disk's perimeter, while profile 2 corresponds only to the critical state, which confines the Meissner phase inside the disk.

Figure 7(a) clearly demonstrates both the reduction in the flux hill's induction when the field jumps from 500 to 600 G, and the sharp increase in the flux hill peak from 430 to 740 G (more than 72%) when the field is switched off (600 G → 0). In the latter case, the flux front penetrates deeper into the sample as the external field is turned off. A similar forward movement of the flux front is also observed in the critical state region (profile 2, figure 7(b)). Magnetic flux with a field of the opposite direction is observed near the edge of the disk due to the demagnetization field, on the magnetic induction distribution after the field is turned off (0 after 600 G).

The right column of figure 7(c-e) shows changes in the induction distribution of the trapped flux hill in the transverse direction under stepwise external field changes. Three cross-sections at different levels are considered: the upper part of the flux hill (Up) – figure 7(c), the maximum flux region (Maximum) – figure 7(d), and the neck separating the flux hill from the critical state belt region (Down) – figure 7(e). The distributions in cross-sections (c) and (d) show that during the flux hill's growth, only the slope with a smaller induction gradient (lower current density) shifts. The flux increases both due to the increase in amplitude at the magnetic field induction maximum and the expansion of the trapped area. This growth makes the peak of the trapped flux more symmetrical. Analyzes of the changes in induction along the flux finger shows that in the region between the critical state belt and the flux finger (cross-section "Down" – figure 7(e)), the induction amplitude decreases slightly, unlike the behavior at the maximum (figure 7(d)).

The evolution of magnetic induction distribution caused by the step change in the field in the SC, described above, with an increase in temperature to 6 K is demonstrated on figure 8. The MO images of induction correspond to field steps $\Delta B_{ext}$: 500 after 400 G (c, c*) → 600 G (d, d*) → 0 (e, e*). The change in magnetic field is caused by the same sequence of actions as shown in figure 6. Vertical and horizontal arrows in figure 8 indicate the areas where the induction distribution is analyzed (figure 8(f, g, h) – bottom row).

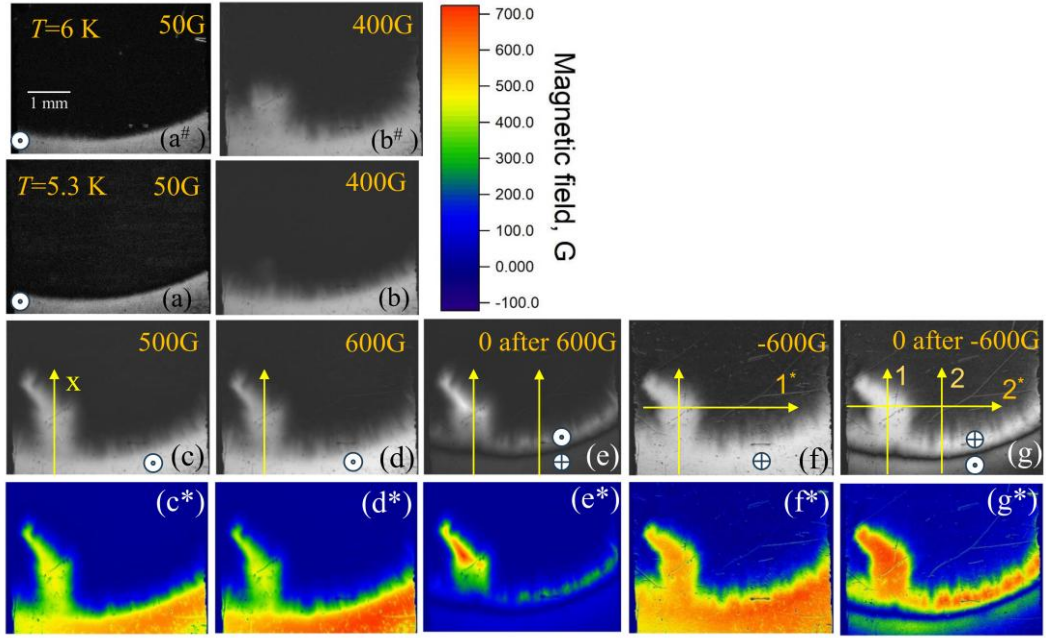

Figure 6. MO images of magnetic flux penetration by steps $\Delta B_{ext}$: 0 → 50 G – (a); 100 G → 200 G → 300 G → 400 G – (b); 500 G – (c); 600 G – (d) and 0 after 600 G – (e); -600 G – (f); 0 after -600 G – (g); ⊕ and - opposite directions of the magnetic field in superconductors; the bottom row of figure (c*- g*) is the view of the previous row (c-g) in color; $T$ = 5.3 K; upper row MO images (a#) and (b#) is at $T$ = 6 K for comparison with (a) and (b) at $T$ = 5.3 K; NbTi alloy after heat treatment. The profiles shown in figure 7 are analyzed along the arrows in the pictures ((c) – (g)).

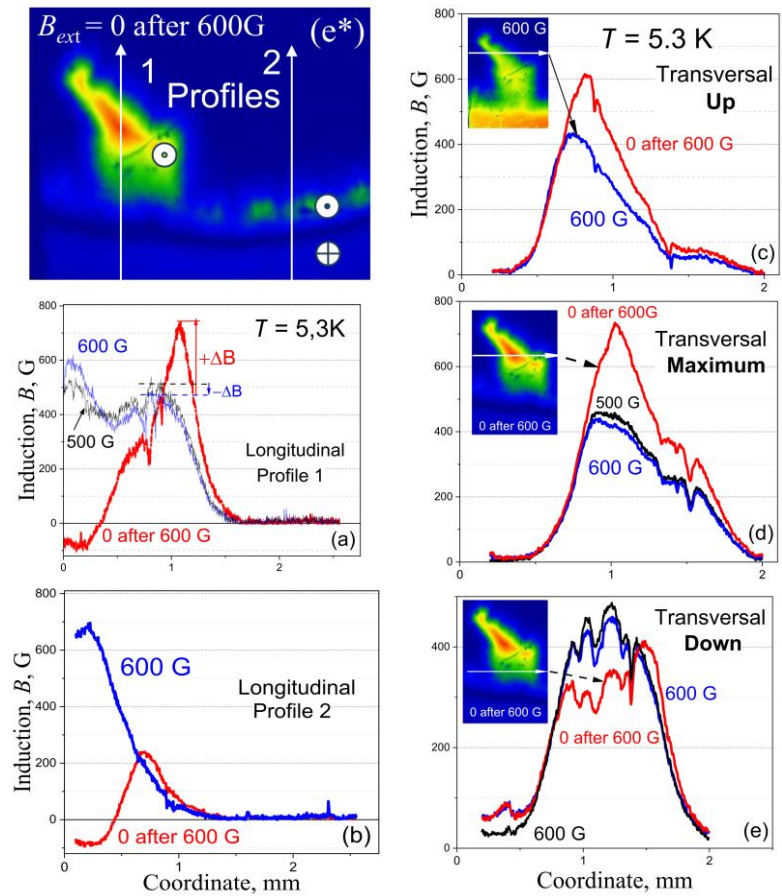

Figure 7. (e*) – MO image of the flux trapped at $B_{ext}$ = 0 after 600 G (figure 6(e*)); arrows 1 and 2 are longitudinal directions of cross section of induction (profiles 1 is in finger range of avalanche and profile 2 is in critical state range); ⊕ and - opposite directions of the magnetic field in superconductors. The distribution of induction along the directions: arrow 1 in avalanche finger for 500 G, 600 G and $B_{ext}$ = 0 after 600 G (a) and arrow 2 for 600 G and $B_{ext}$ = 0 after 600 G (b). Analysis of transversal distributions of induction across the avalanche flux finger in the various cross sections shows in the inserts in the figures: (c) – "Up", (d) – "Maximum" and (e) – "Down". NbTi alloy after heat treatment; $T$ = 5.3 K.

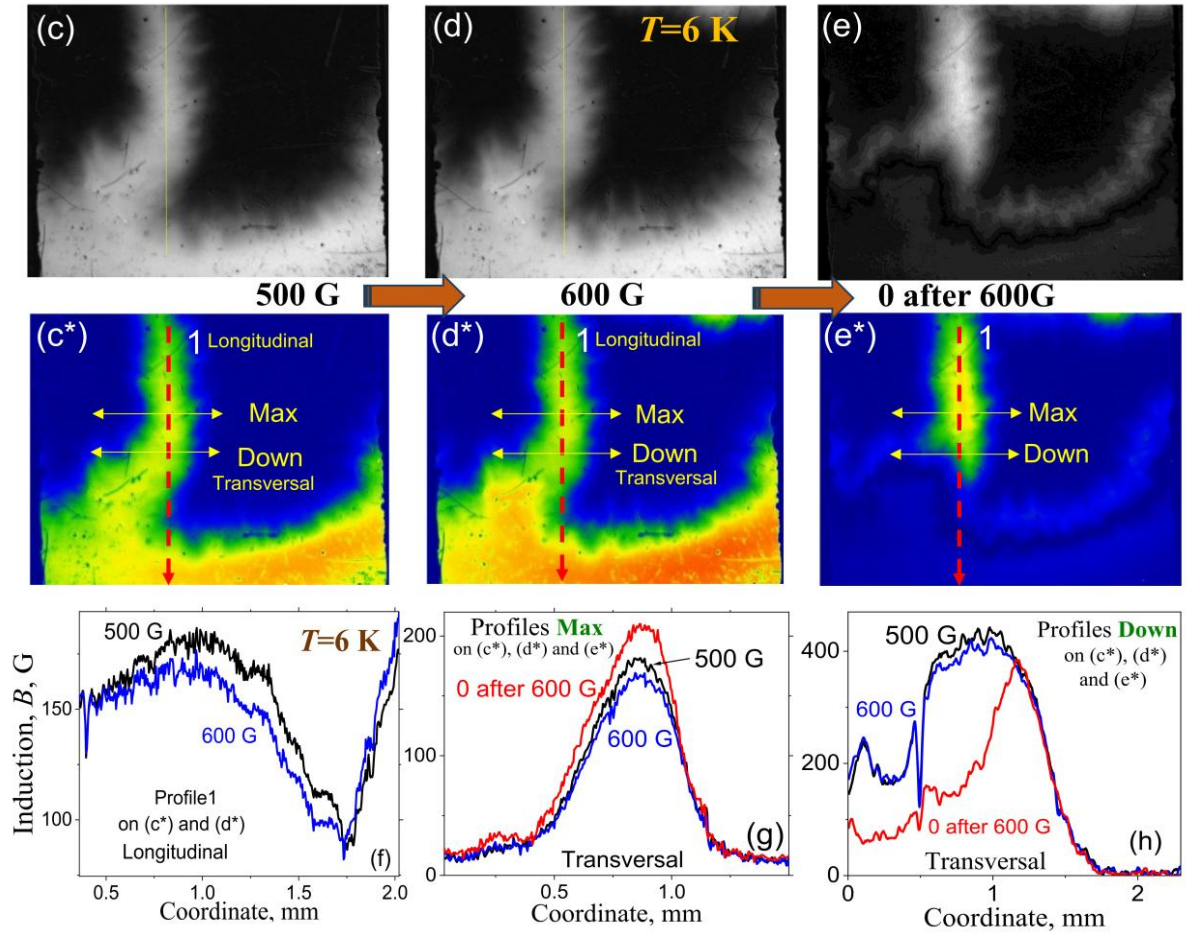

Figure 8. Upper line (c, d, e) – MO images of the magnetic field (500 G, 600 G and 0 after 600 G) penetration and (c*, d*, e*) – the same images in color. Bottom line (f, g and h) presents the distribution of induction along of: (f) – longitudinal direction 1 – for 500 (c*) and 600 G (d*), transversal direction in the positions: (g) – profiles "Max" and (h) – profiles "Down" for 500 G (c), 600 G (d) and 0 after 600 G (e); $T = 6$ K; NbTi alloy after heat treatment.

As seen in figure 8(c*), at a temperature of 6 K, more flux enters the gate into the trapped flux finger, and it penetrates deeper into the superconductor compared to the behavior at 5.3 K (see figure 6(c*) for comparison). The behavior of the flux hill in the finger in both the longitudinal and transverse directions during subsequent field steps from 500 to 600 G (figure 8(d*)) and when the external field is switched off at 600 G (figure 8(e*)) is similar to what was previously described: pulses of increasing or decreasing magnetic fields result in a corresponding decrease or increase in the magnitude of the trapped flux (see the induction distributions in figure 8(f and g). The only difference here is the magnitude of the effect, which has decreased. This is a direct consequence of the reduction in the critical current as the temperature increases. In the "Down" section (figure 8(h)), the observed dynamic changes in the induction distribution are similar to those at 5.3 K.

It is of particular interest to study the dynamic response of trapped flux when magnetic domains with opposite magnetization directions are present. In this case, the magnetic induction distribution contains flux trapped hills with opposing magnetic field orientations. The manipulation of the magnetic field (starting from $B_{ext} = 0$ after +600 G, as shown in figure 9(e, e*) → -600 G (f, f*) → 0 after -600 G (g, g*)) at a temperature of 6 K allowed for the creation of such a magnetic structure in the superconductor. This flux configuration arose when an external field of -600 G was applied to the trapped flux hill formed by the switch-off of the +600 G field (figure 9(e*)). The result is shown in figure 9(f*), where the symbols ⊕ and ⊚, introduced already, indicate opposite directions of the magnetic field in the superconductor. The region of the maximum of the flux captured in the finger at +600 G, which increased as a result of the field being switched off, was entered by a flux finger of opposite polarity (see the region of sections 1 and 2 in the upper part of figure 9(f*)). Although the hill was formed by a +600 G field step (figure 8(d*)), the -600 G field step was not sufficient to fully annihilate the flux hill, as it had increased in amplitude during the switch-off of the +600 G field. As a result, a double-peak magnetic structure, separated by a trough, was formed in the region of profiles 1 and 2, as shown by the black line in figure 9(a).

Figure 9(b) shows the response of these oppositely oriented flux hills to the switch-off of the -600 G field. As can be seen, the dynamic response of the trapped flux hills in both orientations was identical: their heights increased. It is important to note that section 2 passes through the peaks of the hills. At the same time, at the base of the flux hill (section 1), there was no response to the field switch-off (figure 9(c)). This lack of response to dynamic changes at the base of the flux hills was observed in other cases as well.

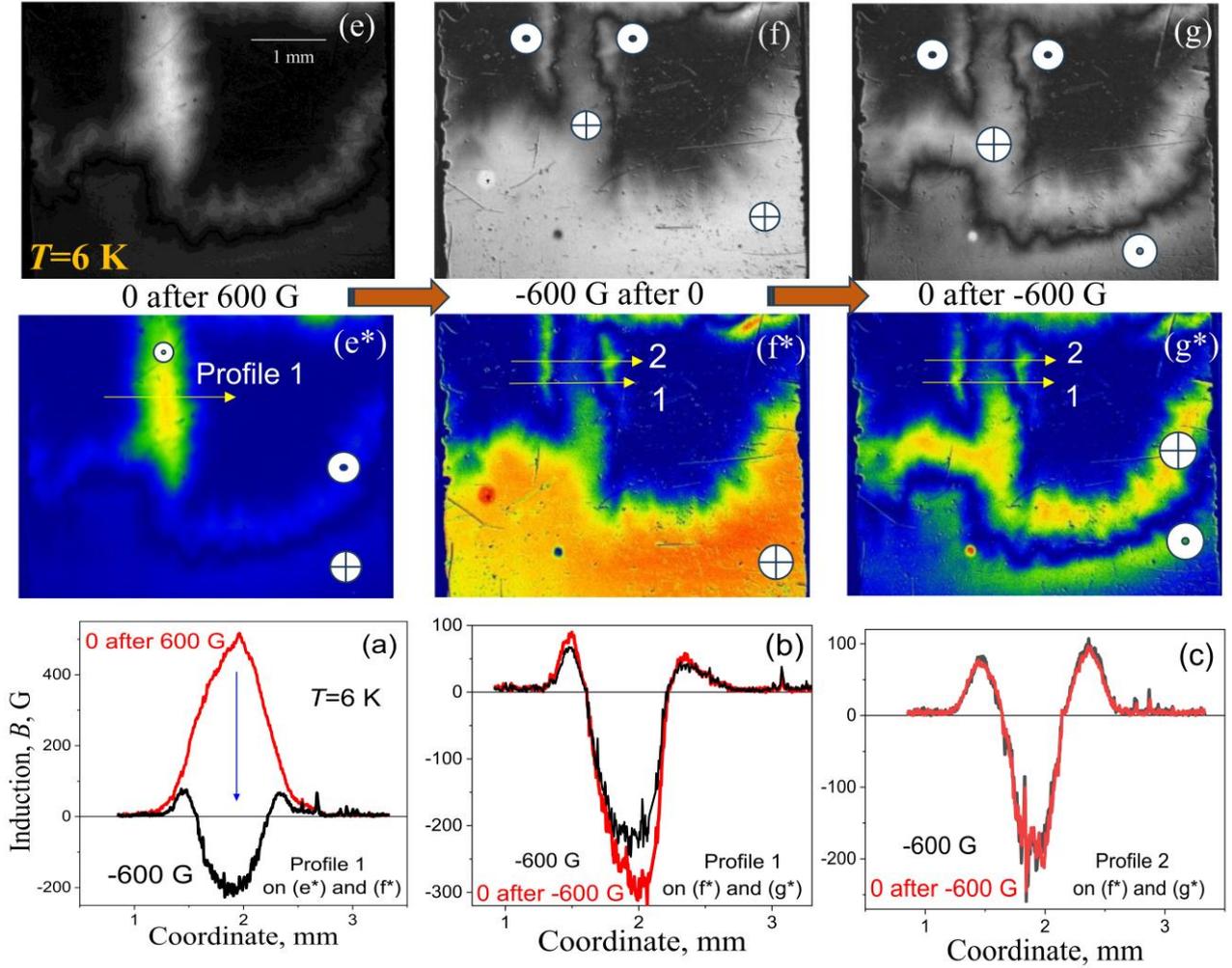

Figure 9. Upper line (e, f, g) – MO images of the magnetic field penetration (0 after 600 G, -600 G after 0, and 0 after -600 G) and (e*, f*, g*) – the same images in color; ⊕ and ⊙ – opposite directions of the magnetic flux in superconductors. Bottom line (a, b and c) presents the distribution of induction along of transversal direction: (a) – profile 1 – for 0 after 600 G (e*) and -600 G after 0 (f*); (b) – profile 1 and (c) – profile 2 for -600 G after 0 and 0 after -600 G (e*); $T$ = 6 K; NbTi alloy after heat treatment.

The growth of a flux hill in an Nb single crystal superconductor during the switch-off of the magnetic field was seemingly observed in Ref. [41], although the authors did not comment on this phenomenon. In that study, when an external field of 800 G was applied, magnetic flux entered through a gate (similarly to that presented in figure 6(b, c)), formed by the twin boundary of the single crystal (figure 4 in Ref. [41]). Upon turning off the field, the amplitude and width of the trapped flux hill increased in the MO image.

*3.3 Mechanism of response of the critical state with trapped flux to stepwise changes of the external magnetic field*

The decrease or increase in the height of the trapped flux hill during the stepwise rise or fall of the external field (figure 7(a, d)) becomes clear when analyzing the sequence of induction responses from current loops in the superconductor, as a reaction to changes in the external field.

Figure 10(a) shows a segment of the penetrating magnetic field, including a circular critical state region and a trapped flux hill that entered through a gate during the rise of the external field to 500 G (figure 8(c)) and the subsequent decrease in the flux hill when the field was further stepped up to 600 G (figure 8(d)). The arrow indicates the direction along which the induction profile was analyzed (profile 1). The trapped flux hill is clearly visible in the induction distribution (figure 10(b)).

The mechanism behind the dynamic changes in the trapped flux hill induction (figure 10(a, b)) becomes clearer when the disk in a magnetic field is represented as a current circuit diagram (figure 10(c)), shown with a ring-shaped shielding current, $J_{scr}$, around the perimeter of the Meissner state. Inside this ring is an oval closed current loop, $J_{trap}$, representing the trapped flux. It should be noted that the ring currents in the disk flow in opposite directions [40]. The two inductively coupled ring currents in figure 10(c) are similar to the current configuration in a superconducting ring that occurs after a flux avalanche enters a hole [42].

During the stepwise change of the external field, $\pm\Delta B_{ext}$, a corresponding shielding current impulse, $\pm\Delta J_{scr}$, is induced along the disk's edge. The magnetic moment, $M_{scr}$, of the shielding current opposes the magnetic moment, $M_{trap}$, of the trapped flux hill inside the disk. Therefore, the induced current impulse, $\pm\Delta J_{trap}$, driven by the jump in $\Delta M_{scr}$, magnetically decreases or increases the trapped flux. As a result, after the relaxation of the superconductor's mixed state, the trapped flux hill becomes larger or smaller than before the external field step.

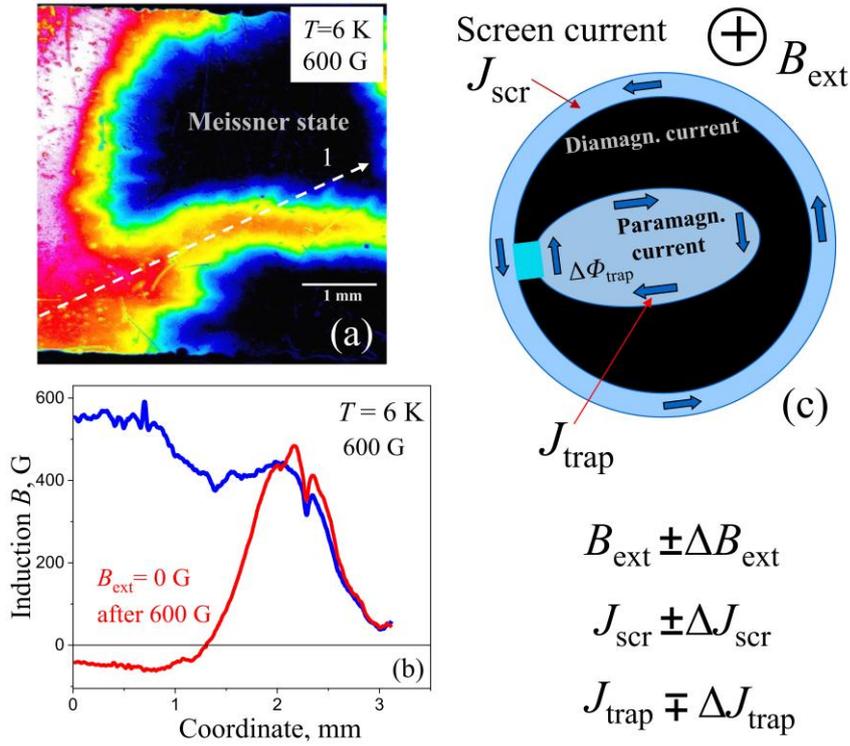

Figure 10. (a) – MO images of the external magnetic field penetration $B_{ext}$ = 600 G at $T$ = 6 K. (b) – The distribution of magnetic induction along direction of arrow 1 on (a). (c) – Schema of the current contours in the superconducting disk after the flux avalanche with a trapped flux $\Delta\Phi_{trap}$ in the external field $B_{ext}$; screening current $\Delta J_{scr}$ and current around of the trapped flux $\Delta J_{trap}$ are response in SC disc, stimulated by step $\Delta B_{ext}$ of external magnetic field.

***Relevance for applications.*** The dynamic behavior of the magnetic flux hill (figure 8(f-h)) discussed here is a dissipative process, leading to heating of the superconducting magnet. Regular pulse-periodic effects on magnet, such as those occurring in electric power generators [4-6], may result in oscillatory movement of the flux front of trapped flux and, consequently, to an increase in the temperature of the magnet. This dissipative dynamic process is an additional source of heating to the radiative heat emitted from the generator armature coils [6] and leading to the decay of trapped flux of field poles.

Considering the form factor of a superconducting magnet in the form of a disk, for example, in a generator, near the edge of the disk along the perimeter, there is always a ring domain with a flux of the opposite sign to the trapped one. In the presence of a boundary separating domains with opposite magnetization directions, these flux movements can lead to vortex-antivortex annihilation accompanied by the release of heat. This process is also dissipative and, together with the dynamics of the flux hill, can result in additional heat generation, particularly in disk-shaped generator magnets.

The observed dynamic changes in the induction distribution of the superconducting permanent magnet under external electromagnetic influences should also be considered when designing and utilizing such magnets in critical applications, such as electromagnetic turnouts and guideways in high-temperature superconducting Maglev systems [43].

***Antiflux avalanches***. The most critical effect on the state of a superconducting disk with trapped flux (cooled in a field of ±600 G and then switched off) occurred when the opposite magnetic field (∓600 G) was applied. Figure 11 shows three instances (columns: (a), (b), (c); (d), (e), (f); and (g), (h), (i)) of these sequential actions at various temperatures. It should be emphasized that, unlike the previous MO images, the magnetic field induction in the superconductor is shown using crossed polaroids, where white areas correspond to one orientation of trapped flux, and black areas indicate the opposite orientation.

The top row of figure 11(a, d, g) shows regions of the disk with trapped flux (field cooled in ±600 G, followed by switching off the field) for three temperatures: (a) – 6.5 K, (d) – 6.2 K, and (g) – 6.1 K. In the first two images, the trapped flux is seen as a black spot, while in the third, it is white. The second row represents the result of dynamic transformations after the application of an oppositely oriented field. Previously trapped flux regions are now disrupted by avalanches of flux with opposite polarity (antiflux avalanches). Flux dendrites with opposite polarity (anti-flux dendrites) have been observed already in superconducting $MgB_2$ rings [46].

The antiflux fingers are highly distorted due to inhomogeneous pinning and take on a dendritic structure. Branching occurs when the flux hot spot reflects off inhomogeneities where the magnetization currents either vanish or change direction [44, 45]. The result of a simulation of thermomagnetic avalanches in a superconducting film with randomly distributed edge defects [45], kindly provided by Prof. I. Aranson [47], is shown in figure 11(j). Green color indicates the magnetic field, and pink represents the temperature. The simulated avalanche pattern demonstrates similarities to the structures observed in our experiment.

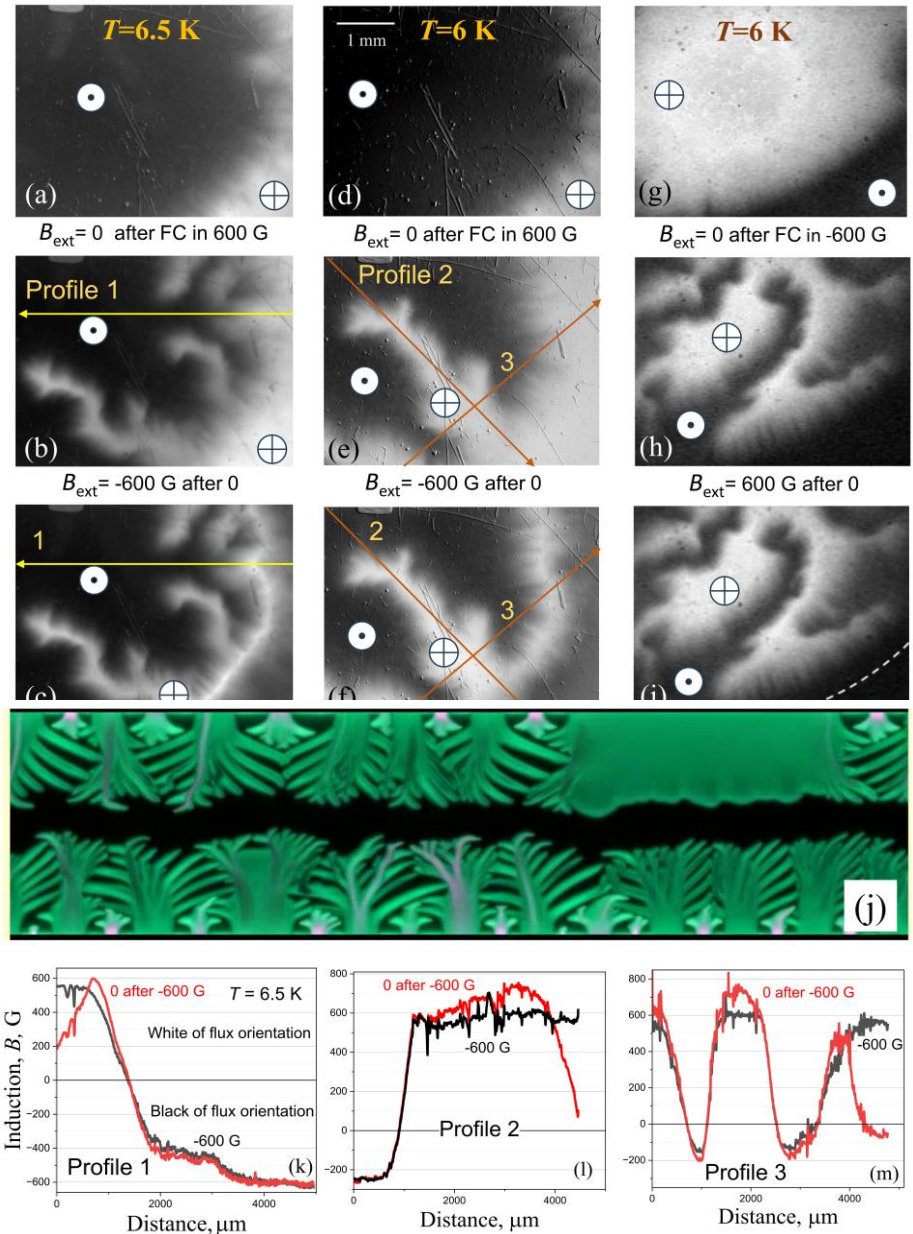

Figure 11. MO images illustrate the occurrence of antiflux avalanches with varying structural variants in NbTi disc. A brighter and the dark image intensity corresponds to the opposite directions of magnetic field. In the top row, the critical state of a superconductor with flux trapped in FC mode ((a), (d) – $B_{ext}$ = 600 G, (g) – -600 G) is depicted, with the external field turned off ($B_{ext}$ = 0) at temperatures: (a) – 6.5 K, (d) and (g) – 6 K. Outside the sample there is a stray field with a direction opposite to the direction of the trapped flux field. The second row (b), (e), (h) shows cases of MO images recording avalanches that occurred after switching on a 600 G field with a direction opposite to the direction of the field of the trapped flux (a), (d), (g), respectively. The third row – (c), (f), (i) illustrates the transformation of the induction distribution resulting from turning off the external field. The arrows in this context indicate the directions in which the analysis of the induction distributions, shown in (k)–(m), was conducted; in the view (i) at the bottom right, the dotted section of the arc shows approximately the edge of the disk. (j) – one fragment of simulated flux penetration (avalanches) in SC film the finger undergoes multiple branching, giving rise to the characteristic flux dendrites [44, 45]. A larger number of edge defect is present. Green color corresponds to magnetic field and pink to the temperature.

The third row of figure 11(c, f and i) shows the trapped flux after the external field has been switched off. Changes in the induction near the disk edge due to the demagnetization field are clearly visible. In figure 11(i), the edge of the superconducting disk is marked with a dashed line.

It was of particular interest to analyze the transformation of the trapped flux structure in highly inhomogeneous induction regions (second and third rows of figure 11) after switching off the field. Arrows with numbers on figure 11(b, c, e and f) indicate the directions along which this analysis of magnetic induction distribution changes was performed. The bottom row (figure 11(k, l, m)) presents these distributions. As can be seen, the induction distribution response is consistent everywhere, both in the antiflux fingers and in the critical state region along the disk's perimeter: the absolute flux magnitude increases due to the dynamic field effects. In terms of image brightness, the lighter areas become brighter, and the darker areas become darker. For example, the induction increased by 23% in the dendrite-like antiflux finger (figure 11(l), profile 2). In profile 3 (figure 11(m)), each induction extremum clearly shows an increase in amplitude after the field was switched off (0 after 600 G).To clarify the dynamic changes observed in the images, we conducted induction measurements using a Hall sensor at the center of the disk during remagnetization. Flux jumps were recorded using an inductive sensor – a coil placed around the disk. Figure 12(a) shows the hysteresis loop of $B_{surf}(B_{ext})$. Particular attention was given to the area of flux entry during the reversal of the external field and to magnetic flux creep (figure 12(b)). The magnetic field rate was reduced to minimize the possibility of thermomagnetic avalanches and to detect the flux creep behavior. However, even with the reduced field rate, small magnetic flux jumps still occurred (figure 12(b)). These are typical thermomagnetic avalanches with millisecond timescales [48]. The structure of one such jump (jump 1), with a duration of $\Delta t_{av} = 1.7$ ms, is shown in the inset of figure 12(b).

This remagnetization area, immediately after field reversal, exhibited a rich variety of effects. A complex pattern of field and current was revealed in a superconducting ring [49], where three concentric current loops were detected.

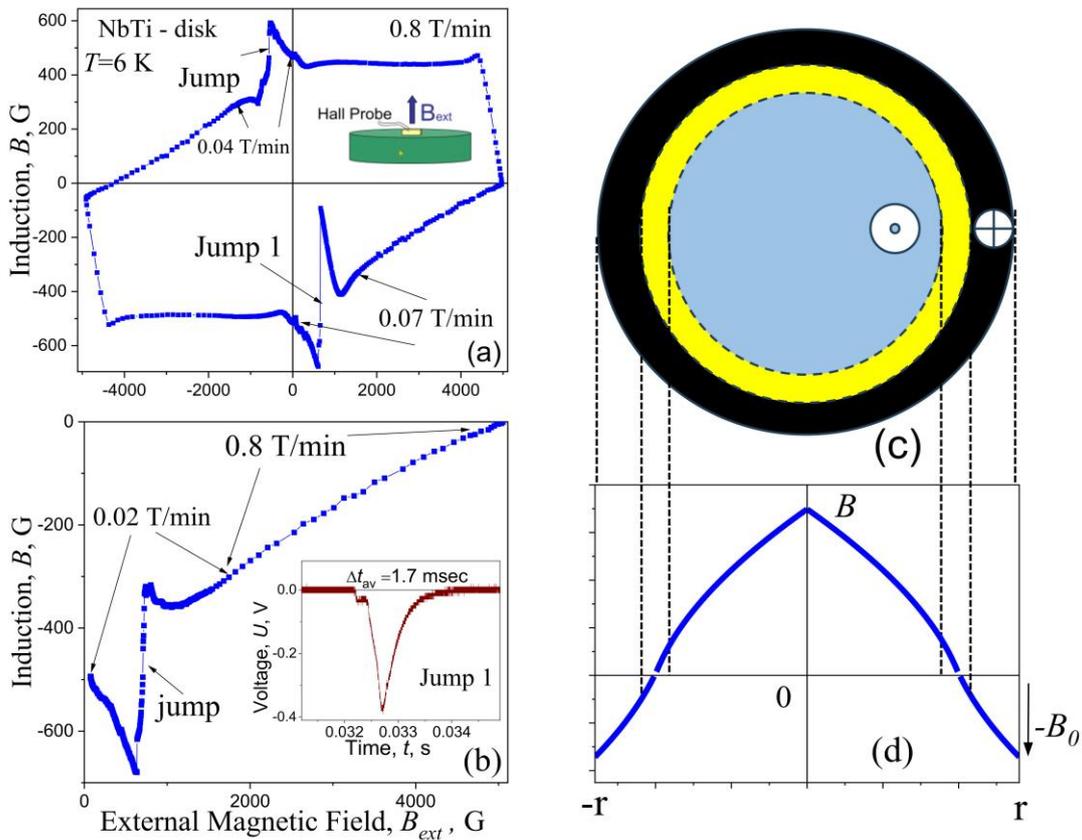

Figure 12. (a) Main frame: hysteresis loop of remagnetization $B(B_{ext})$ at $dB_{ext}/dt = 0.8$ T/min; right inset – geometry of experiment; $B_{ext}$ is the induction of external magnetic field in the center of disc. (b) – the detailed behavior of induction at slow change of external field $dB_{ext}/dt = 0.02$ T/min after changing the direction of the magnetic field; inset – time dependence of the voltage $U_{coil}(t)$, registered during the flux jump 1; $\Delta t_{av}$ is the duration of flux jump; $T = 6$ K. (c) – Scheme of a disk with a trapped flux in the center (blue color) and with a negative field direction (black ring) in the external field $B_0$; (d) – calculated induction distribution $B(x)$ in SC in Kim-Anderson model of critical current (from Ref. [50]).

What could be the physical reason for the significant "destructive" impact of the opposite field on the critical state of a disk with trapped flux? This instability in the critical state during remagnetization is characteristic of samples with various shapes: cylinders, plates [50, 51], films [52], and others. In a disk, the instability is more pronounced due to the large demagnetization field and the energetically unfavorable, strongly curved structure (see figure 5) of the external field lines.

Additionally, significant magnetic flux creep in the remagnetization region (the yellow ring in figure 12(c)) likely occurs due to vortex-antivortex annihilation, which is accompanied by localized heat generation and, consequently, a reduction in flux pinning in that region.

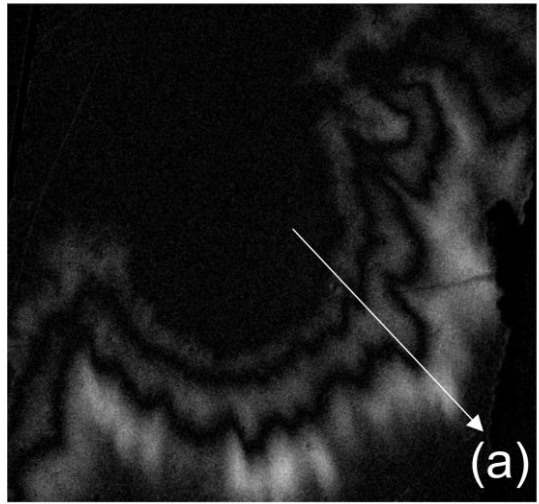
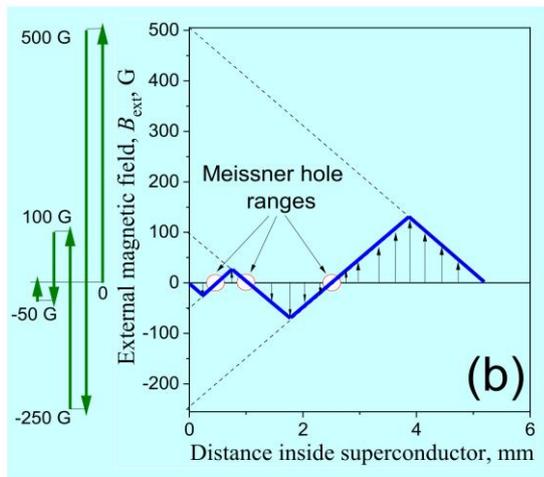
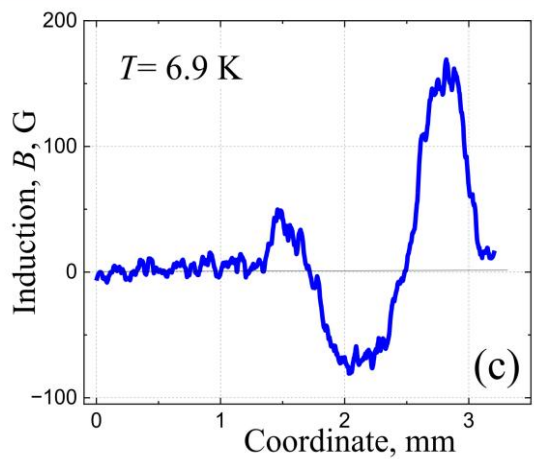

Figure 13. (a) – MO view of induction pattern after remagnetization on scheme presented on (b); in this pattern, a brighter image corresponds to a higher normal component of induction, and a darker image corresponds to a lower one. Arrow shows the direction of magnetic induction analysis; (b) - the magnetization reversal of a superconductor with decreasing amplitude and a change in magnetic field direction at each step $\Delta B_{ext}$: 0 → 500 G → -250 G → 100 G → -50 G → 0. (c) – magnetic induction profile along the arrow on (a); $T$ = 6.9 K, $B_{ext}$ = 0 G.

*Alternating magnetization reversal and analysis of pinning homogeneity in the material.* Alternating magnetization reversal allows the visualization of the pinning structure throughout almost the entire volume of the superconductor. This structure occurs after processes such as plastic deformation during extrusion or after annealing.

The magnetization reversal of the superconductor (see figure 13(a)) was performed by reducing the field strength by approximately half and changing its direction at each step $\Delta B_{ext}$, according to the scheme shown in figure 13(b): ZFC procedure than 0 → $\Delta B_{ext}$ = 500 G → -250 G → 100 G → -50 G → 0. At the final stage, after reducing the field to zero, an image was taken. The induction pattern after remagnetization, according to the scheme in figure 13(b), for the extruded NbTi alloy at $T$ = 6.9 K, is shown in figure 13(a). In these images, brighter areas correspond to a higher normal component of induction, while darker areas represent a lower component. Here, the "negative" and "positive" fields of the trapped flux are separated by dark lines, where vortices and antivortices annihilate. The boundary between the two flux directions is highly curved and covered with small, sharp roughness.

As seen in figure 13(a), the shape and structure of the dark lines vary depending on their location on the disk, reflecting changes in the structure of the pinning centers. Therefore, if the maximum field $B_{ext}$ in the remagnetization scheme is chosen to be close to the full penetration field (similar to the penetration shown in figure 1(b)), the dark lines at the remagnetization fronts in the proposed scheme can cover the entire disk surface.

Figure 13(c) shows the magnetic induction profile in the direction indicated by the arrow in figure 13(a). The large and small sharp roughness along the dark lines at the trapped flux front in NbTi produce a "noisy" curve, making it difficult to analyze the detailed structure. It is worth noting that in single crystals of high-temperature superconductors ($YBa_2Cu_3O_{7-x}$ [53] and $MgB_2$ [54]), in a certain temperature range, the profiles of the normal component of induction near the remagnetization front reveal significant changes, particularly sharp kinks. Unfortunately, it was not possible to identify characteristic features in the structure of these regions within the available temperature range for our study. Nevertheless, the striped domain structure with irregular boundary roughness may still characterize the degree of pinning center homogeneity. Scaling analysis of the induction profiles at the Meissner level and the roughness index can help qualitatively characterize the heterogeneity of the pinning structure.

*3.4 Analysis of flux profiles at the Meissner level and behind the front: roughness exponents and Harsdorf dimension*

The study of flux front structure has been conducted in many works (see, for example, [34, 35]). Their analysis demonstrated that the behavior of the magnetic flux front penetrating a superconductor is similar to phenomena such as paper combustion, fluid infiltration through porous media, crystal growth, etc., i.e. it has a fractal nature. However, as pointed out by the authors of Ref. [34], there is a significant difference between these phenomena and flux dynamics in superconductors. In the first three cases (combustion, growth, infiltration), the structure behind the front is fixed, whereas in superconductors, the magnetic flux distribution can change in the region behind the front.

This underscores the necessity of studying not only the flux front's shape at the Meissner level but also the distribution of magnetic induction in the penetrated region. To achieve this, lines of constant induction were determined within the flux-penetrated area for a given external field.

According to current understanding (see, for example, Refs [34, 35, 55]), the structure of the penetrating flux front is determined by the structure of the pinning centers and the stochastic nature of vortex dynamics. The influence of stochasticity on the trapped flux structure is characterized by the degree of reproducibility in the induction patterns. To assess the role of stochasticity, we repeatedly captured flux structures under the same experimental conditions. It is worth paying attention that while the roughness exponents are consistent within experimental error, the induction distribution (i.e., the shape of the isocontours) varies slightly.

Figure 14 presents two examples of such structures for a disk made from extruded material (a, b) and for the annealed alloy (d, e). In both cases, large-scale roughness is well-reproduced, whereas fine-scale roughness appears differently structured. This suggests that pinning on structural defects is stronger than on fine-grained inclusions, such as normal-phase α-titanium particles, distributed throughout the material. The stochastic nature of vortex dynamics may manifest primarily on these α-titanium particles.

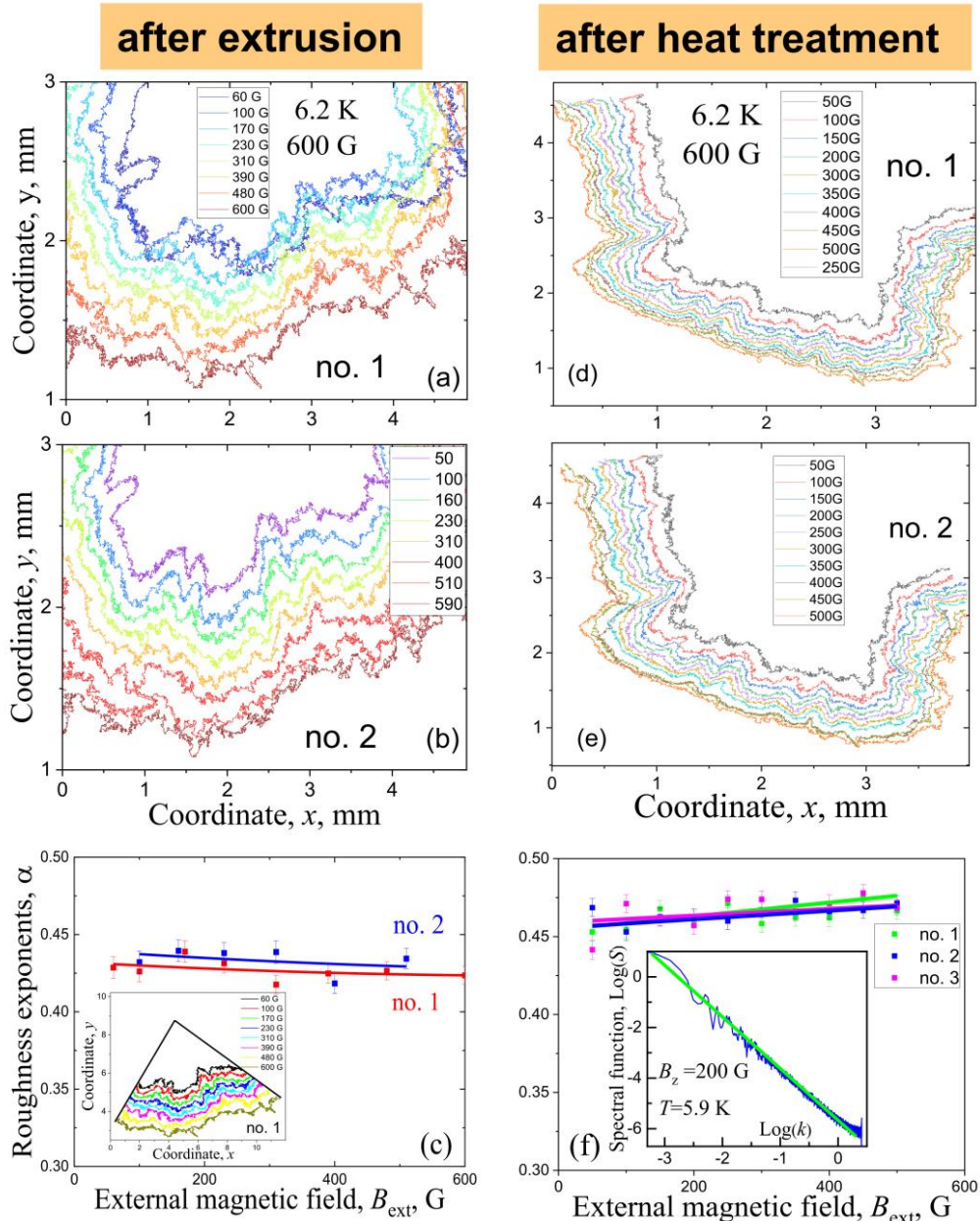

Figure 14. Lines of the front sections of the induction profile at various levels for two measurements, no. 1 and no. 2, repeated under the same conditions: (a, b) - SC after extrusion and (d, e) - SC after heat treatment, $T = 6.2$ K; (c, f) - magnetic field dependence of roughness parameter $\alpha$ for two measurements no. 1 and no. 2 after extrusion (c) and three measurements after heat treatment (f); insert to (c) shows the spectral function $S(k)$ on a double logarithmic scale and its approximation by a linear dependence; $B_{ext} = 600$ G; NbTi disc.

A comparison of the images of large-scale flux front roughness structures in the left (a, b) and right (d, e) columns of figure 14 shows that annealing smooths out the large roughness. This could be due to the relaxation during annealing of large-scale defects, such as grain boundaries, which were formed in the crystalline structure as a result of severe deformation during extrusion.

It should be noted that using the MO method, the analysis of large-scale roughness in hard superconductors after extrusion may provide insights into the plastic flow of the material under various deformation methods, particularly in the case of screw extrusion [56].

On the other hand, the fine-scale roughness structure does not repeat itself. This could be caused by pinning on α-titanium grains, which are about $10 \times 10$ μm$^2$ in size – an order of magnitude smaller than the magnetic domain in figure 4(d). In this case, reproducing the fine-scale induction distribution formed by vortex bundles on α-titanium particles seems unlikely, especially since the measurement error in determination of the induction level exceeds the size of the fine pinning particles.

To analyze the shape of the curves representing the flux front, we used the Fast Fourier Transform (FFT) method and constructed the power function $S(k)$. In Ref. [34], it was shown that the roughness exponent obtained with this method fully coincides with the results obtained using other methods. A plot of this function in double logarithmic scale, for one of the flux fronts, is shown in the inset of figure 14(f). From the graph, it is evident that within a certain range of $k$ values, the function $\log S(k)$ can be approximated by a linear relationship, i.e., $S(k)$ behaves as a power function. The roughness exponent $α$ of the flux front is determined by the standard procedure [34, 35, 57] from the slope of $\log S(k)$ versus $\log(k)$. A unified criterion – minimizing the standard error – was used when determining the slope for all curves.

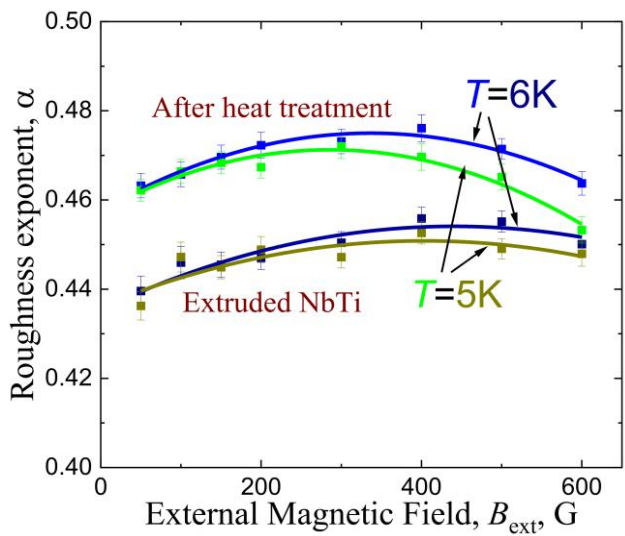

Figure 15. Magnetic field dependence of roughness parameter $α$ for flux front on Meissner level at 5 and 6 K; NbTi alloy after extrusion and after heat treatment.

The dependence of the roughness exponent $α$ as a function of the magnetic field is shown in figure 14(c, f). The field-dependence statistics indicate that, within an error margin of 10–15%, the $α(B_{ext})$ points fit linear trends. However, in the case of the extruded alloy, this roughness exponent slightly decreases with increasing field (figure 14(c)), whereas in the annealed alloy, it shows a slight upward trend (figure 14(f)). The decrease in roughness with increasing magnetic field in the extruded alloy may be due to the fact that at higher fields, the isocontours are closer to the edge of the sample, leading to their alignment by the current flow near the edge, as suggested by the authors of Ref. [34]. In the annealed sample, an increasing number of pinning centers secure more vortices as the field increases, and this effect outweighs the alignment process.

The roughness exponent in the extruded material is approximately 8% lower than in the annealed alloy. This correlates well with the idea that annealing leads to the relaxation of defects and stresses in the deformed, extruded structure, reducing the displacement size of the large-scale roughness at the flux front. The pinning centers become smaller and more evenly distributed throughout the volume, which increases the small-scale roughness of the flux front and thus raises the roughness exponent. Figure 15 shows the magnetic field dependence of the roughness parameter $α$ for the flux front at the Meissner level at temperatures of 5 and 6 K for NbTi alloy after extrusion and after heat treatment. Analysis of the data shows that, at least in this temperature range, temperature has only a slight effect on the roughness exponent. As the temperature rises, the roughness exponent increases slightly, both before and after heat treatment so, heat treatment has a more significant impact and as a result of annealing, the roughness exponent increases. As noted above, annealing significantly enhances the critical current, indicating a change in the pinning structure. We believe that the change in the roughness exponent is also related to the modification of the pinning structure. During annealing, the number of pinning centers increases, and they become smaller and more uniformly distributed throughout the superconductor. This results in a more homogeneous distribution of pinning centers, which eliminates the large-scale granularity present before annealing.

The dependence of the flux front roughness exponent on the external field (figure 15) is non-monotonic: initially, it increases with the field, but then it starts to decrease. At the initial stage, the front becomes aligned along the sample's edge, and as the field increases, the roughness slightly rises due to the presence of pinning centers. The subsequent decrease in the roughness exponent with further field increase could be due to the greater energy released during vortex motion, which causes the smaller pinning centers to be bypassed, making the flux front structure less rough. At higher external field strengths, more vortices enter the superconductor, and they move faster. This factor could be significant, as, unlike the authors in Refs. [34, 35], who studied thin films, our experiment involves a bulk sample.

As shown in figure 15, annealing led to an increase in the roughness exponent in agreement with the observation that the flux front structure at the Meissner level became finer as a result of annealing compared to the extruded material. The decrease in the roughness exponent, $α$, in a strong field reflects the fact that the size and number of large-scale roughness features increase with increasing external magnetic field. This behavior of the large-scale roughness is also evident in figure 2 (f and h).

Next, we examined the induction isocontours at various levels for trapped flux in both the extruded sample and the annealed one. After the external field induction reached 600 G, the field was switched off, and the field distribution in the trapped flux was analyzed (figure 16(a) at $T$ = 6.2 K and figure 16(d) at $T$ = 5.3 K). Figure 16(b and e) show the induction isocontours in the extruded material and after annealing, respectively. The spacing between the isocontours on the slopes reveals that in both cases, the outer slope (near the disk's edge) of the trapped flux is steeper than the inner slope. Another obvious fact is that the inner slope of the induction profile has more pronounced large-scale roughness. This again suggests that the magnetic field induction is more aligned along the edge of the sample.

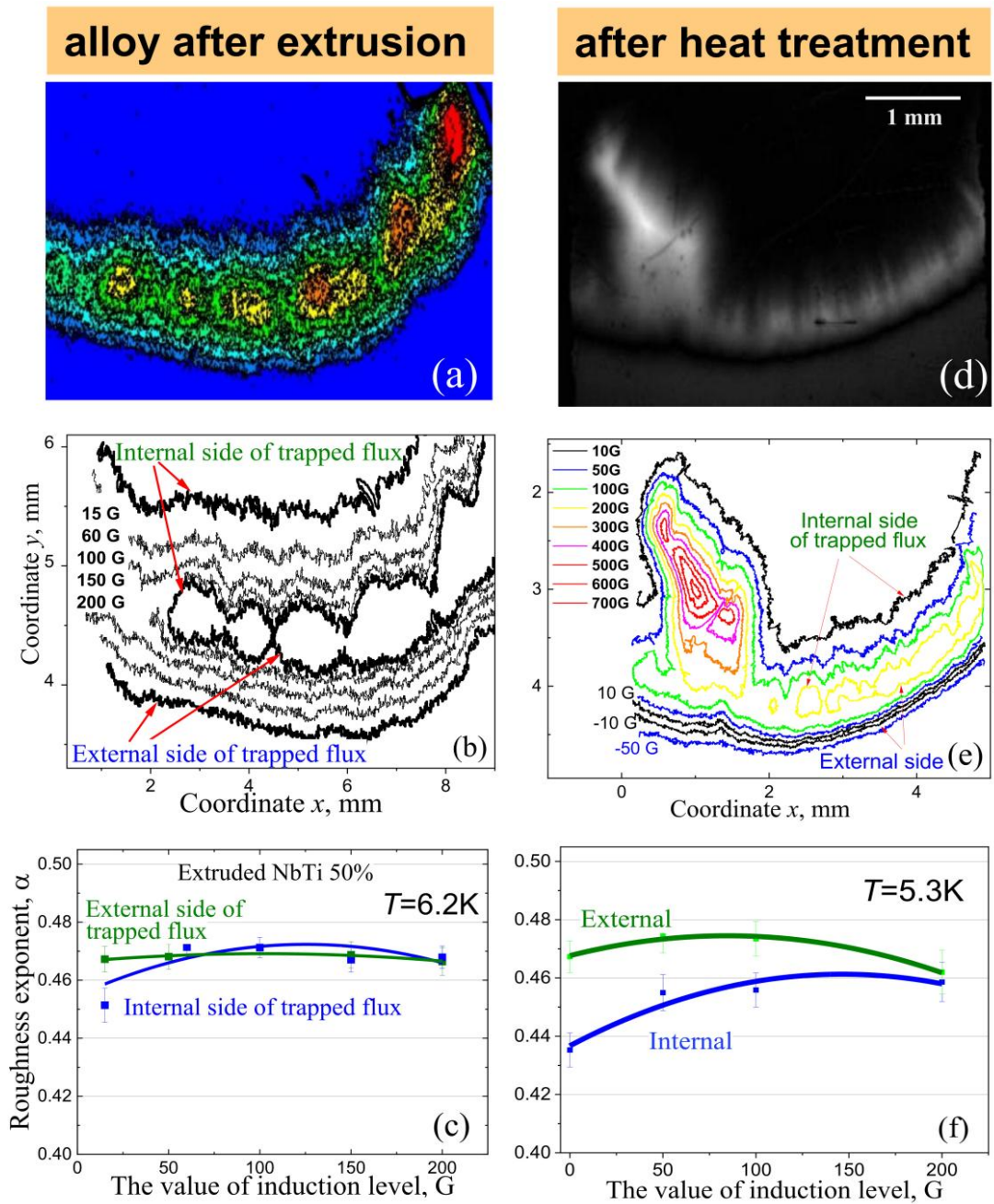

Figure 16. Data for NbTi disc alloy after extrusion at $T$ = 6.2 K – left column and after heat treatment at $T$ = 5.3 K – right column: trapped flux views and lines of cross sections of the induction profile at various value levels (a, d), the induction contours into on various levels (b, e), and roughness exponent $\alpha$ of induction contours for external and internal sides of trapped flux presented on (c, f).

The roughness exponent, $\alpha$, for the induction isocontours on both the outer and inner sides of the trapped flux region was calculated at various induction levels (figure 16(c and f)). Both roughness exponents increase slightly as the induction level rises, and they converge as they approach the maximum, which is expected.

The obtained roughness exponent values lie in the range of 0.435–0.475, while the Hausdorff dimension is between 1.525 and 1.565. If we accept, as many authors suggest, that the flux front has a fractal (self-affine) nature [34, 35, 56-59], we must recognize that, in this case, the system follows the dynamic stochastic disorder model. As noted by Kardar, Parisi, and Zhang, the roughness exponent in this model should satisfy the condition $α < 0.5$ (the so-called KPZ model [55]).

The difference between the outer and inner sides of the trapped flux hill (figure 16(b, e)) is determined by the different magnetic history of these material layers. The inner slope represents the frozen flux that entered at a field strength of $H_{ext} ≈ 0 - H_{max}/2$. The outer slope was formed under the pressure of the Lorentz force in a magnetic field significantly larger, $H_{ext} = H_{max}/2 - H_{max}$, and this part of the trapped flux was later expelled from the superconductor due to inter-vortex repulsion.

The magnetic induction (Lorentz force) in a superconductor defines the energy threshold at which pinning centers either become suppressed or remain active. The structural differences in the induction profiles of the inner and outer slopes are influenced by variations in the distribution of active (unsuppressed) pinning centers within these material volumes. Due to the differing magnetic histories of the layers, these distributions vary, ensuring the irreversibility of magnetic properties in hard superconductors.

In figure 16(d), a dark, arc-shaped line is visible in the lower part of the image. This represents the region of magnetization reversal near the edge of the superconductor disk, resulting from the demagnetizing factor. The boundaries of this region are more clearly defined in figure 16(e), between the black induction lines of +10 G and -10 G. The cyclic motion of such a magnetic structure in a permanent superconducting disk magnet, under the influence of periodic impacts in an electric power generator or motor, is a dissipative process driven by vortex/antivortex annihilation.

## 4. Conclusions

Magneto-optical investigations of the behavior of a superconducting Nb50%Ti disk with trapped flux (acting as a permanent magnet) under stepwise changes in an external magnetic field (~ 0.5 T/s rate) were conducted at temperatures between 5 and 7 K. The experiments revealed dynamic changes in the flux trap structure in response to stepwise variations in the external magnetic field. Specifically, each stepwise increase or decrease in the field resulted in a corresponding increase or decrease in the trapped flux. For example, at a temperature of 5 K, a 600 G field step caused the maximum induction of the trapped flux to increase by 40–50%. The magnitude of these changes was influenced by external conditions, highlighting the sensitivity of superconducting magnets to their operating environment.

These forced dynamics in trapped flux, can lead to additional energy dissipation and heating, which may impact the performance and reliability of superconducting magnets in applications. Additionally, stepwise increases in the external magnetic field, oriented opposite to the trapped flux, triggered antiflux avalanches in the superconducting disk.

We conducted a detailed analysis of the flux front structure during flux penetration and trapping for the alloy in two states: after extrusion and after subsequent heat treatment. The analysis of the power function describing the roughness of the flux front, along with the calculation of the Hausdorff fractal dimension, revealed significant insights into the material's pinning properties and the influence of processing conditions.

These findings have practical implications for the design and use of bulk superconducting magnets in applications where they are exposed to alternating (AC) magnetic fields, such as in power generation and magnetic levitation systems. Understanding the dynamic behavior of trapped flux and its effect on energy dissipation and thermal management is critical for optimizing the performance and longevity of these superconducting devices.

## Acknowledgements

We thank Prof. I. S. Aronson for kindly providing the results of simulations of avalanche processes in superconductors.

We also gratefully acknowledge the Defense Forces of Ukraine for enabling us to conduct this research.